\documentclass[aps,prb,preprint,groupedaddress,showpacs,amsmath,amssymb]{revtex4}

\usepackage{graphicx}
\usepackage{dcolumn}
\usepackage{bm}
\usepackage{hyperref}

\begin{document}

\title{Pressure tuning of structural and magnetic transitions in EuAg$_4$As$_2$}

\author{Sergey L. Bud'ko$^1$, Li Xiang$^1$, Chaowei Hu$^2$, Bing Shen$^2$,  Ni Ni$^2$, and  Paul C. Canfield$^1$ }
\affiliation{$^{1}$Ames Laboratory, US DOE, and Department of Physics and Astronomy, Iowa State University, Ames, Iowa 50011, USA}
\affiliation{$^{2}$Department of Physics and Astronomy and California NanoSystems Institute, University of California, Los Angeles, CA 90095, USA}

\date{\today}

\begin{abstract}
We report temperature dependent measurements of ambient pressure specific heat,  magnetic susceptibility, anisotropic resistivity and thermal expansion as well as in-plane resistivity under pressure up to 20.8 kbar on single crystals of EuAg$_4$As$_2$. Based on thermal expansion and  in-plane electrical transport measurements at ambient pressure this compound has two, first order, structural transitions in  80 - 120 K temperature range. Ambient pressure specific heat, magnetization and thermal expansion measurements show a cascade of up to seven transitions between 8 and 16 K associated with the ordering of the Eu$^{2+}$ moments. In-plane electrical transport is able to detect more prominent of these transitions: at 15.5, 9.9, and 8.7 K as well as a weak feature at 11.8 K at ambient pressure. Pressure dependent electrical transport data show that the magnetic transitions  shift to higher temperatures under pressure, as does the upper structural transition, whereas the lower structural transition is suppressed and ultimately vanishes. A jump in resistivity, associated with the upper structural transition, decreases under pressure with an extrapolated disappearance (or a change of sign)  by 30-35 kbar. In the 10 - 15 kbar range a  kink in the pressure dependency of the upper structural transition temperature as well as the high and low temperature in-plane resistivities suggest that a change in the electronic structure may occur in this pressure range. The results are compared with the literature data for SrAg$_4$As$_2$.

\end{abstract}


\maketitle

\section{Introduction}
Since the discovery of high temperature superconductivity in iron-arsenides, \cite{kam06a,kam08a} studies of the physical properties of large families of pnictide-based compounds have attracted a lot of attention. The interplay between different orders, such as the coexistence, competition and/or correlation of  superconductivity, magnetism and structural phase transitions (or nematicity)  in pnictides, \cite{ste11a,chu15a,zap17a,boh17a} made these families a playground for condensed matter physicists, chemists, and materials scientists. Europium - based pnictides occupy a special position in the family since they add local, Eu$^{2+}$ magnetism to the complexity of the observed orders. \cite{zap17a} 

The new, trigonal, layered arsenide, EuAg$_4$As$_2$, and its non-magnetic counterpart, SrAg$_4$As$_2$,  were first synthesized few years ago. \cite{sto12a} In  polycrystalline samples of  EuAg$_4$As$_2$ a divalent nature of europium and antiferromagnetic order at $T_N = 14.9$ K were determined from magnetization and $^{151}$Eu M\"ossbauer measurements. \cite{ger13a} More recently, detailed structural, thermodynamic and transport studies of single crystals of  EuAg$_4$As$_2$ revealed a complex physical picture that includes a structural distortion at 120 K and incommensurate, non-collinear long - range antiferromagnetism developing at 9 K. \cite{she18a} The magnetic transition at $\sim 15$ K \cite{ger13a}, based on detailed recent $^{151}$Eu M\"ossbauer data, \cite{rya19a} was suggested to be a transition from paramagnetic state to an incommensurate sine-modulated long range order.  Motivated by the complexity and interplay of different orders in EuAg$_4$As$_2$ as well as reported curious, non-monotonic pressure dependence of the structural transition in its non-magnetic analogue, SrAg$_4$As$_2$, \cite{she18b} in this work we study tunability of magnetic and structural transitions in EuAg$_4$As$_2$ by hydrostatic pressure via in-plane resistivity measurements.

\section{Experimental details}
Single crystals of EuAg$_4$As$_2$ were grown out of a high temperature melt with an excess of Ag$_2$As. Details of of the synthesis can be found elsewhere. \cite{she18a}  Resulting plate-like single crystals ($c$ - axis perpendicular to the plates) were of several mm in-plane dimensions and over a mm thick.
In order to allow ambient pressure comparison, a detailed set of ambient pressure measurements was done on the particular batch of crystals that was later used for transport measurements under pressure.
Ambient pressure, in-plane and $c$-axis ac ($f = 16$ Hz) resistivity measurements were performed in a standard, linear, four-probe configuration using an ACT option of a Quantum Design Physical Property Measurement System (PPMS). Electrical contacts to the sample were made using Epotek H20 silver epoxy. Magnetization measurements were performed in a Quantum Design MPMS3 magnetometer. Low temperature heat capacity measurements were made using semi-adiabatic thermal relaxation technique as implemented in the heat capacity option of the Quantum Design PPMS.  Anisotropic thermal expansion measurements were carried out using an OFCH copper capacitive dilatometer. \cite{sch06a}

In-plane resistivity measurements under pressure were performed in a hybrid, Be-Cu / Ni-Cr-Al piston - cylinder pressure cell (modified version of the one used in Ref. \onlinecite{bud86a}) in the temperature environment provided by a PPMS instrument.  A 40 : 60 mixture of light mineral oil and n-pentane was used as a pressure-transmitting medium.This medium solidifies at room temperature in the pressure range of 30 - 40 kbar, \cite{bud86a,kim11a,tor15a} which is above the maximum pressure in this work. Elemental Pb was used as a low temperature pressure gauge. \cite{eil81a} The measurements were performed both on increase and decrease of pressure and the results are reversible. It has been known  \cite{its64a,its67a,bec76a,fuj80a,tho84a} that in piston-cylinder pressure cells high temperature pressures are different from low temperature pressures and the temperature dependence of this pressure difference is non-trivial.  Given that the temperature/pressure relation for this specific cell/pressure medium combination has not been established, here we simply use the Pb gauge pressure value. This may give rise to  pressure differences, at higher temperatures, of up to 3 kbar.

\section{Results}
\subsection{Ambient Pressure}
\begin{figure}
\begin{center}
\includegraphics[angle=0,width=140mm]{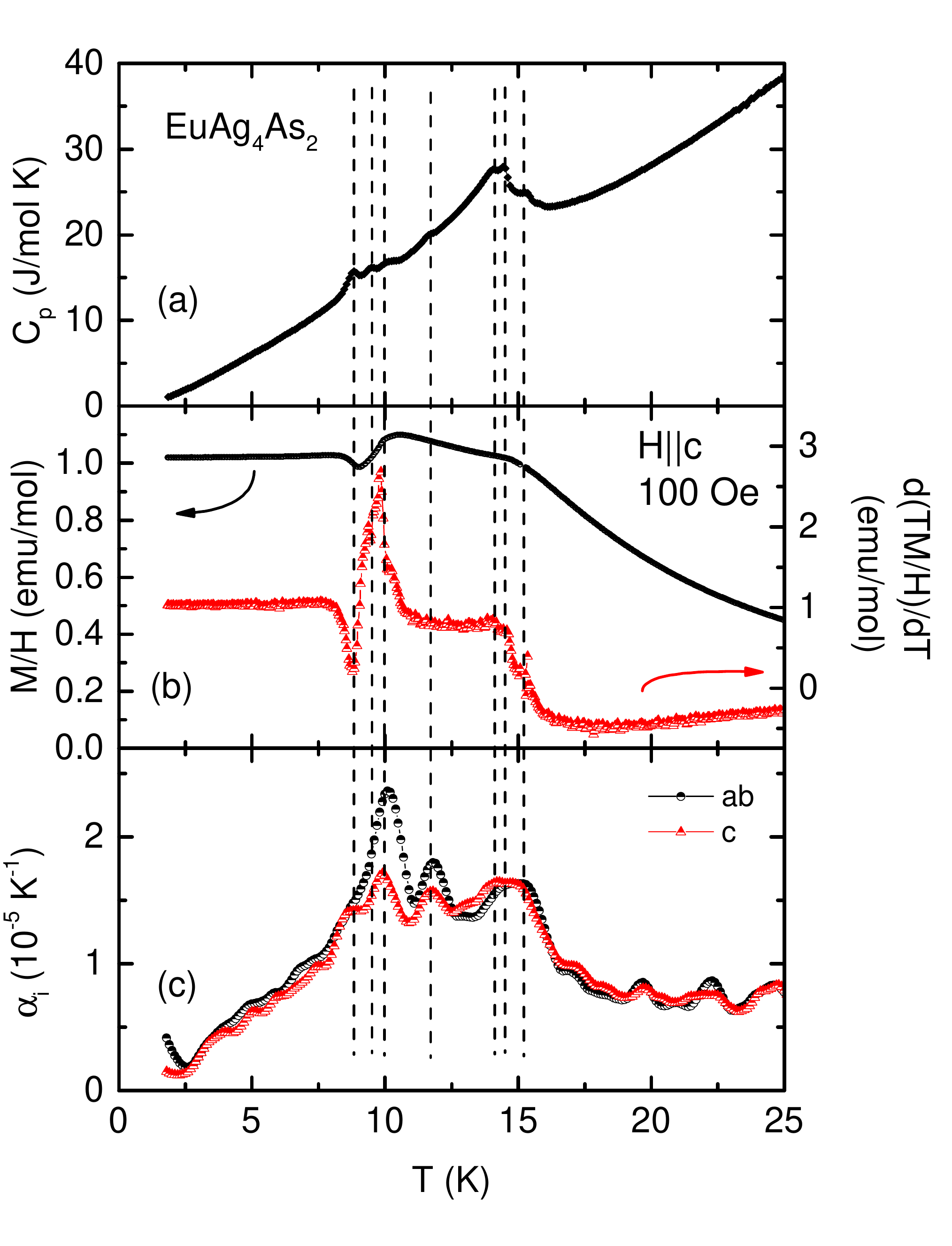}
\end{center}
\caption{(color online)  Low temperature part of (a)  heat capacity, (b)  magnetic susceptibility, $M/H$ measured in magnetic field $H = 100$ Oe applied parallel to the $c$ - axis and the temperature derivative,  $d (T~ M/H)/dT$,\cite{fis62a} and (c)  anisotropic thermal expansion coefficients, $\alpha_i$. Heat capacity data were taken on cooling, magnetization and thermal expansion - on warming. Vertical lines mark transitions as seen in low temperature heat capacity. } \label{F1}
\end{figure}

Low temperature heat capacity data are shown in Fig. \ref{F1}(a). The data are complex and suggest up to seven transitions (marked with vertical dashed lines),  occur in EuAg$_4$As$_2$ below  16 K as the Eu$^{2+}$ moments order. Low temperature, low field magnetic susceptibility, $M/H$ and its derivative in a form of $d(T~M/H) / dT$ \cite{fis62a} have anomalies at similar temperatures (Fig. \ref{F1}(b)), although the one at $\sim 11.8$ K is not discernible, at least for  $H \| c$. This identification of possible transition temperatures show fair agreement with the anisotropic thermal expansion data shown in Fig. \ref{F1}(c). Altogether, three different thermodynamic measurements indicate as many as seven, closely spaced low temperature transitions. Such a density of transitions is remarkable, but not unprecedented; CeSb has seven transitions between 8 and 18 K in zero applied magnetic field and even and even larger number of additional transitions in applied fields. \cite{wie00a} 

Anisotropic thermal expansion (Fig. \ref{FN4}) serves as a thermodynamic probe of the phase transitions (the structural ones in particular). Indeed, both structural phase transitions are clearly seen in the thermal expansion $[L_i(T) - L_i(1.8 \textrm{K}]/L_i(1.8 \textrm{K})$ data, where $L_i$ is the sample's length, either along the $c$-axis or in $ab$ - plane. Whereas the lattice change is large and in the same direction for both lattice parameters at the $T_1$ transition, the response smaller and  is anisotropic for the $T_2$ transition. Below 17 K, up to five of the seven transitions detected in specific heat and magnetization data are also seen in the thermal expansion coefficients (Fig. \ref{F1}(c)). The changes of the thermal expansion coefficients through the magnetic transitions are qualitatively similar for both orientations. There are no obvious anomalies in the anisotropic thermal expansion data at and above $\sim 225$ K, thus the anomaly detected in resistivity data in this temperature range is either associated with a very broad and subtle structural transition, or is some artifact pertinent to transport measurements. Additional studies are required to address this issue.

\begin{figure}
\begin{center}
\includegraphics[angle=0,width=120mm]{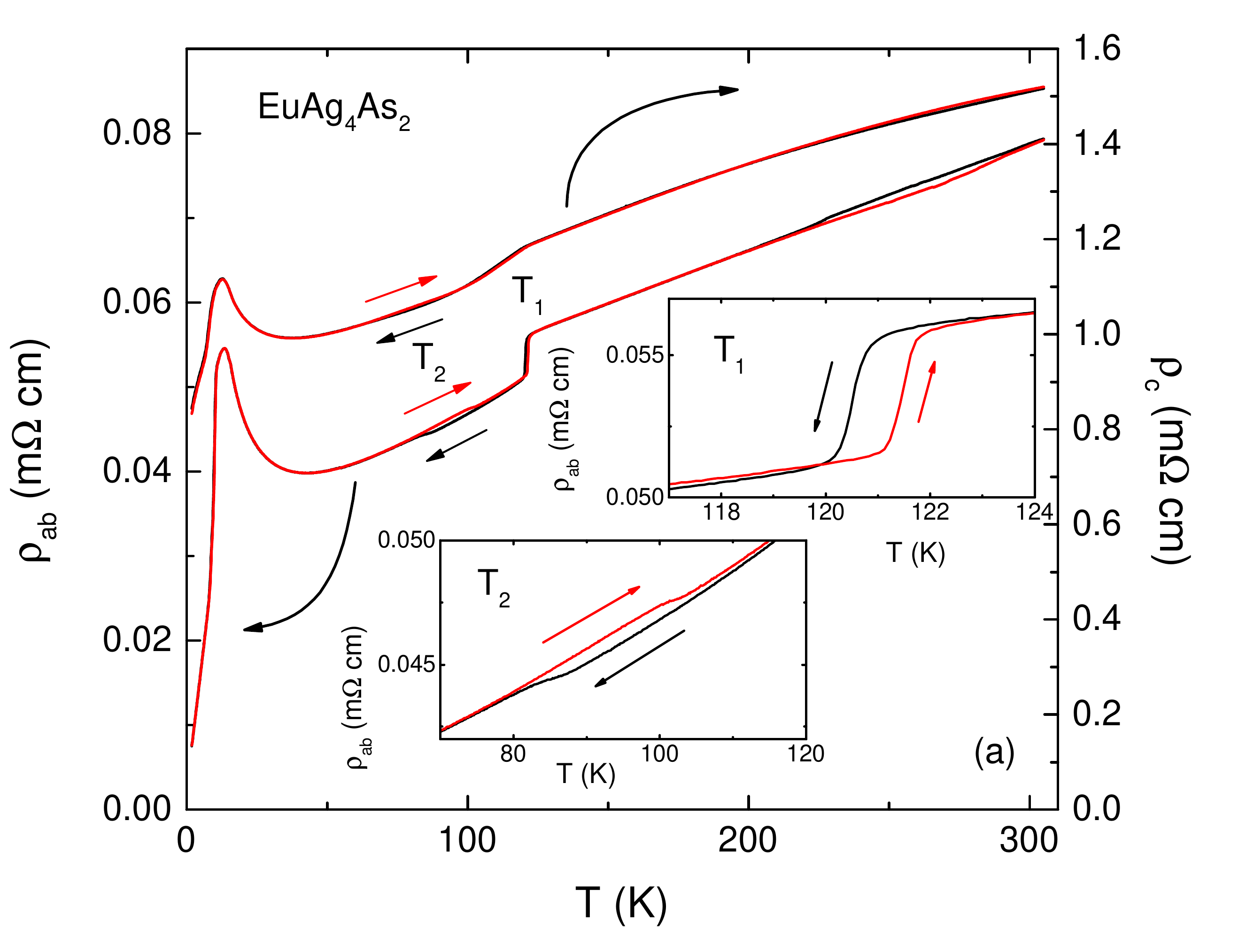}
\includegraphics[angle=0,width=120mm]{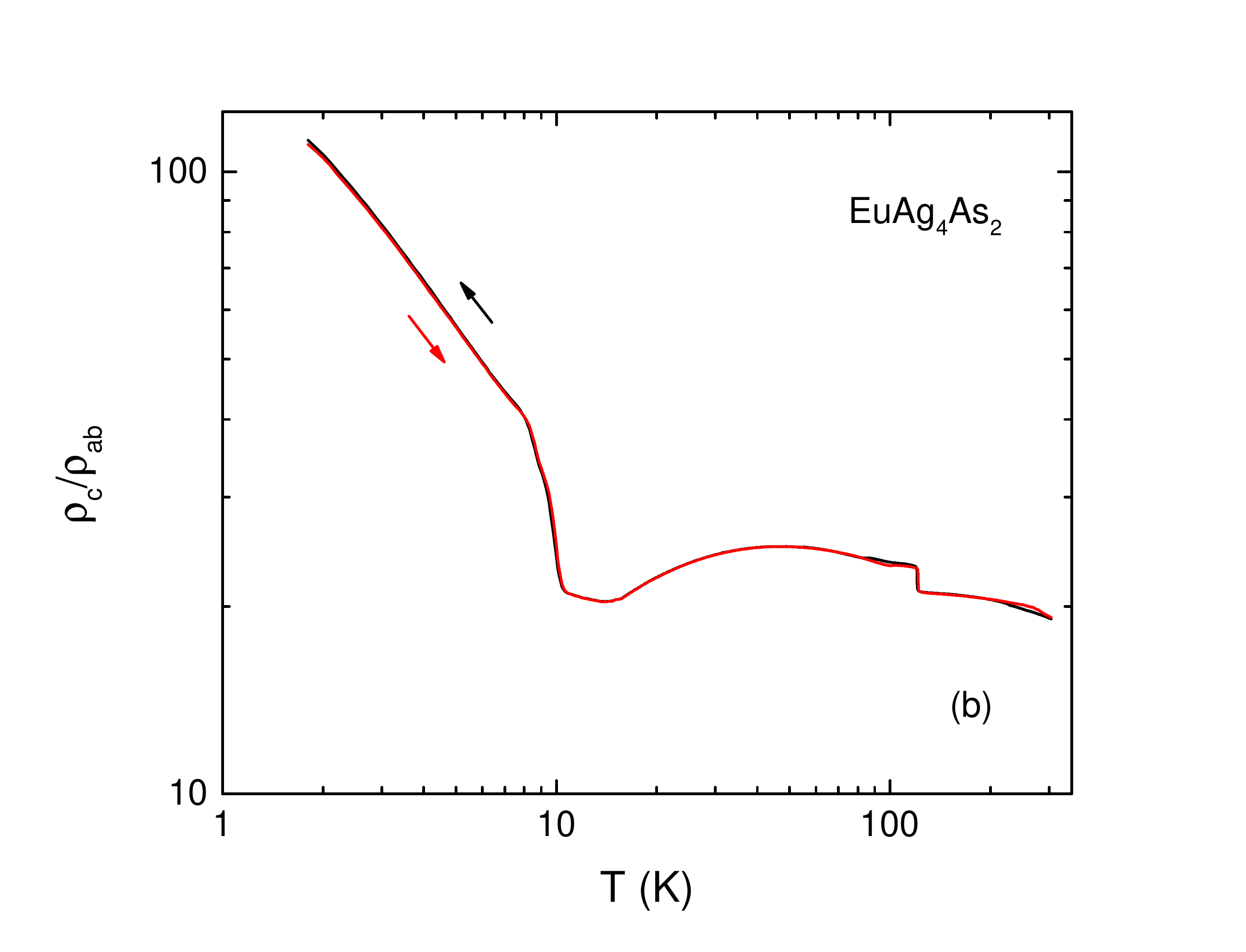}
\end{center}
\caption{(color online) (a) Anisotropic ambient pressure resistivity of  EuAg$_4$As$_2$ measured on cooling and on warming. Insets: hystereses of the structural phase transitions as seen in the in-plane resistivity. (b) Resistivity anisotropy of  EuAg$_4$As$_2$ from the data taken on cooling (black) and on warming (red).} \label{F2}
\end{figure}

Anisotropic ambient pressure resistivity is shown in Fig. \ref{F2}(a). There are several points of note. The in-plane residual resistivity ratio, $RRR = \rho(300 \textrm{K}) / \rho(2 \textrm{K}) \approx 10$ in this work, is almost a factor of 2 higher than that in Ref. \onlinecite{she18a} that suggests better crystallinity or fewer defects and impurities. The structural phase transition  $T_1$ is sharper and $\sim 10$ K higher than reported. \cite{she18a}  Moreover, another,  hysteretic, possibly structural transition $T_2$ can be detected in the 85 - 100 K range. These two transitions are also seen, although somewhat less clearly, in the $c$- axis resistivity data. There might also be another, very broad and hysteretic, transition above $\sim 225$ K (Fig. \ref{F2}(a)). 

At room temperature the anisotropy of resistivity, $\rho_c/\rho_{ab}$ is about 20 (Fig. \ref{F2}(b)). This value changes significantly through the structural and magnetic transitions thus reflecting anisotropic contribution of related changes in scattering and electronic structure to the electronic transport properties. The value of  $\rho_c/\rho_{ab}$ reaches $\gtrsim 100$ at the base temperature. This may be primary associated with the large residual resistivity, $\rho_0$, term (smaller RRR value) found for the $\rho_c$ data. As expected, the behavior of  $\rho_c/\rho_{ab}$ is very similar, on log-log scale, for the data taken on cooling and on warming with slight differences seen for the $T_2$ transition and above $\sim 225$ K. An $\sim 1$ K difference is present in the data for the $T_1$ transition, even if it is not clearly seen on the large scale log-log plot of Fig. \ref{F2}(b).  Both in-plane and $c$ - axis resistivities have pronounced upturn above magnetic ordering. This upturn is probably due to scattering on magnetic fluctuations.

\begin{figure}
\begin{center}
\includegraphics[angle=0,width=140mm]{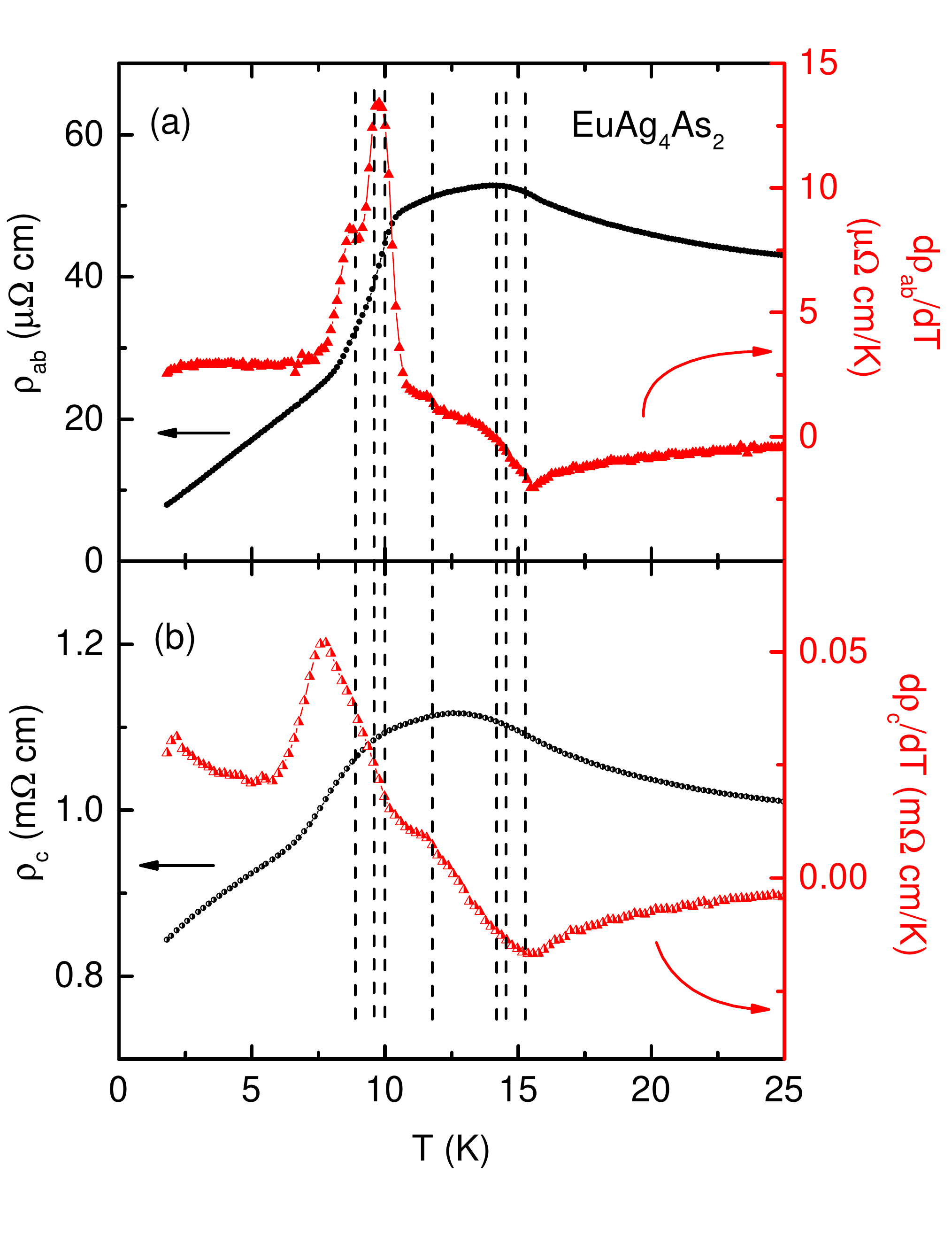}
\end{center}
\caption{(color online)  Low temperature part of (a) in- plane, (b) out-of- plane resistivity  and their temperature derivatives. Vertical lines mark transitions as seen in low temperature heat capacity in Fig. \ref{F1}(a). } \label{F3}
\end{figure}

The low temperature part of the in-plane resistivities, $\rho_{ab}$ and $\rho_{c}$, together with their respective temperature derivatives, $d \rho_{ab} / dT$ and $d \rho_{c} / dT$ are shown in Fig. \ref{F3}(a),(b). There are three clear features in the  $d \rho_{ab} / dT$, suggesting three  transitions at 15.5 K, 9.9 K, and 8.7 K, \cite{fis68a} as well as another, possibly spin reorientation transition, at about 11.7 K . The corresponding feature in $\rho_{ab}(T)$ and its derivative is rather subtle. The out-of-plane resistivity data give the same transition temperatures if for two lower transitions different criterion, the middle point of the increase in $d \rho_{c} / dT$ \cite{tau16a} is utilized. 

The comparison of low temperature thermodynamic (Fig. \ref{F1}) and transport (Fig. \ref{F3}) data leads to several conclusions. The feature in $d \rho_{c} / dT$ at $\sim 7.5$ K does not have its counterpart in any of the thermodynamic measurements presented in Fig. \ref{F1} so it does not seem to be related to a thermodynamic phase transition. Although the low temperature transitions seen in $d \rho_{ab} / dT$ are only a subset of the transitions occurring in EuAg$_4$As$_2$, they do correlate with the primary features and span of the transitions detected and can serve as a caliper of how the magnetism in EuAg$_4$As$_2$ responds to pressure. 

Based on the above, measurements of the in-plane resistivity allow for tracking of two high temperature structural transitions as well as of a subset of the low temperature magnetic transitions. In the following we present the $\rho_{ab}(T)$ measurements under pressure in order to provide an initial mapping of the $P - T$ phase diagram for EuAg$_4$As$_2$.

\subsection{Resistivity under Pressure}
\begin{figure}
\begin{center}
\includegraphics[angle=0,width=120mm]{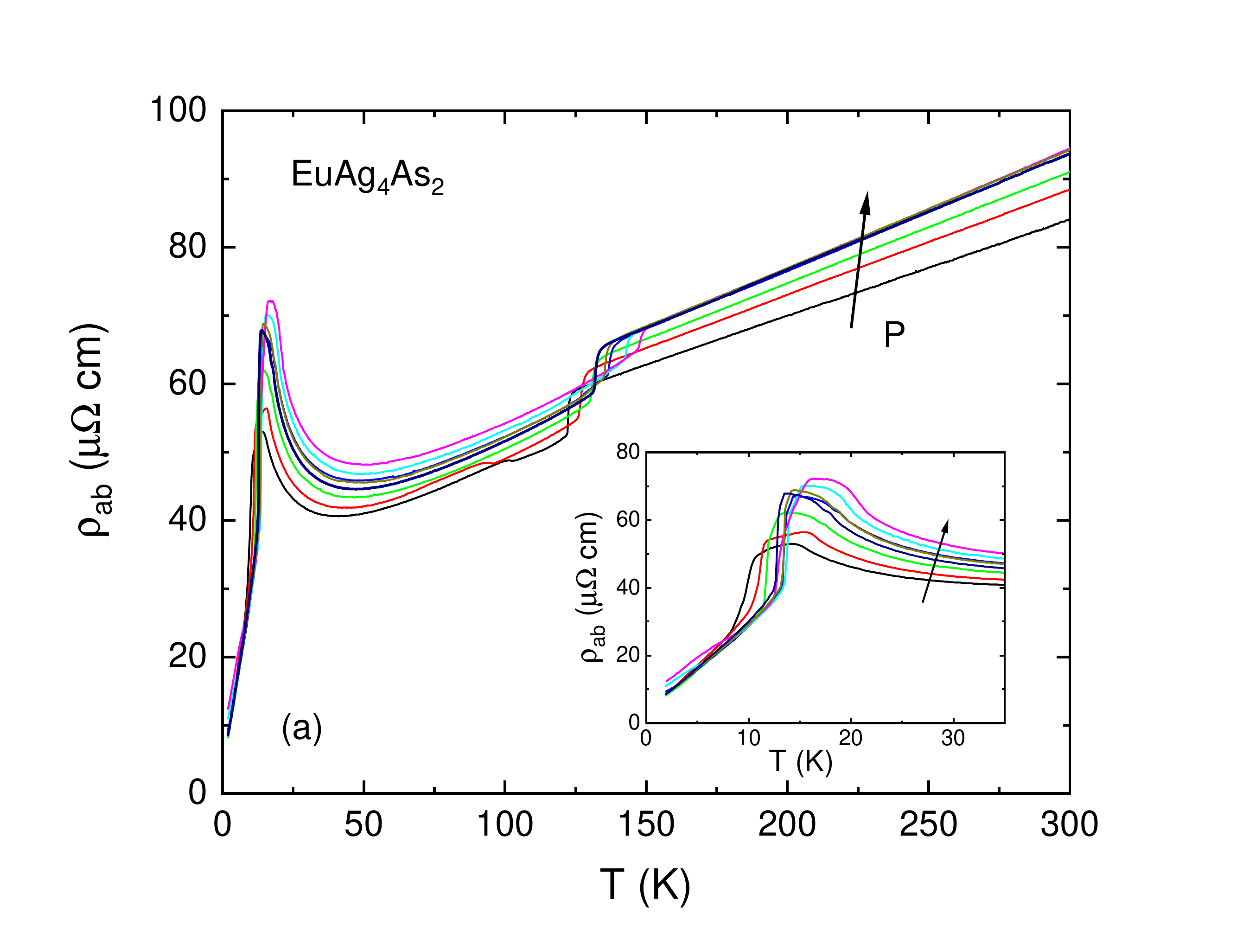}
\includegraphics[angle=0,width=120mm]{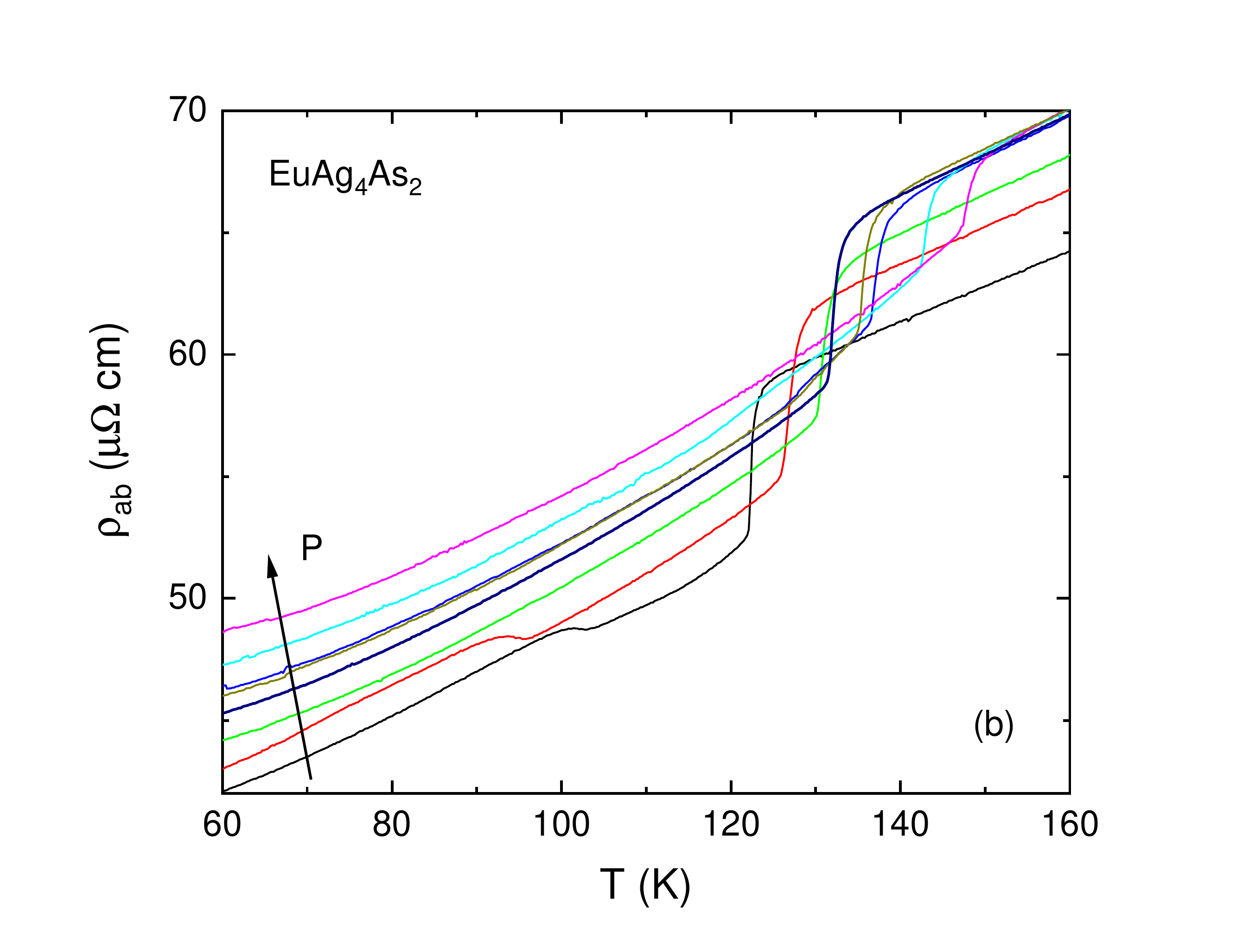}
\end{center}
\caption{(color online) (a) Temperature - dependent resistivity of EuAg$_4$As$_2$ measured at different pressures (subset of the data is shown). Inset: enlarged region of magnetic transitions. (b) Enlarged region of structural transitions. Arrows show the direction of the pressure increase. The low temperature values of pressure are: 0.02, 4.0, 9.4, 10.5, 13.6, 15.1, 18.2, and 20.8 kbar.  Data taken on warming are shown.} \label{F4}
\end{figure}

The evolution of the in-plane resistivity under pressure up to 20.8 kbar is shown in  Fig. \ref{F4}. With applied pressure, the upper structural transition moves up, as do the lower temperature, magnetic transitions.   The lower structural transition appears to move down under pressure (Fif. \ref{F4}(b)) and soon becomes undetectable. As pressure increases, the upturn above the magnetic transitions becomes more pronounced and, away from  the phase transition temperatures, the resistivity generally increases too.

\begin{figure}
\begin{center}
\includegraphics[angle=0,width=140mm]{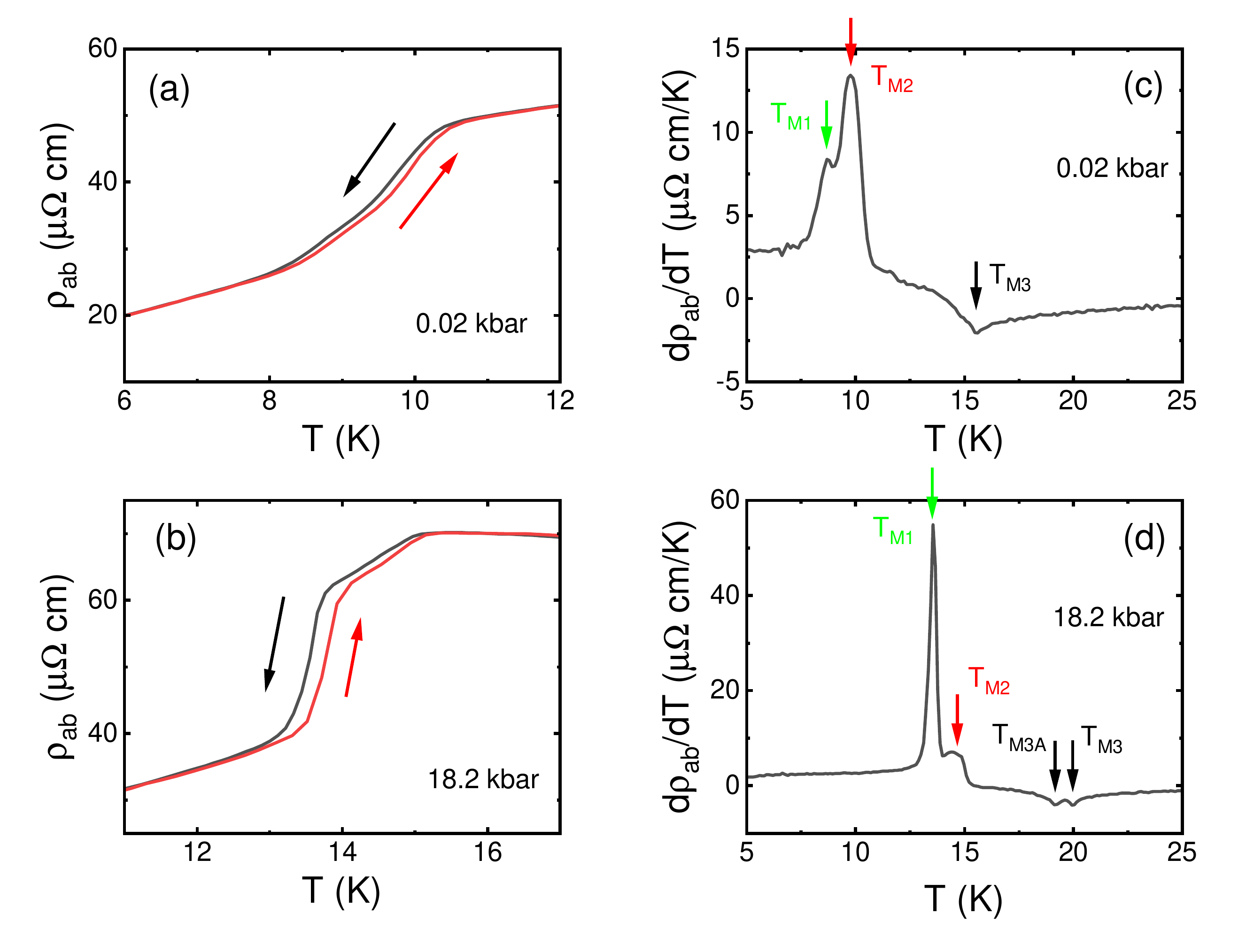}
\end{center}
\caption{(color online) (a), (b) Low temperature resistivity measured on cooling and warming; (c), (d) low temperature derivatives $d\rho_{ab}/dT$ (for measurements on cooling) at two representative pressures, 0.02 kbar and 18.2 kbar. Arrows mark transitions. Note that upper transition splits in two under pressure.} \label{F5}
\end{figure}

\begin{figure}
\begin{center}
\includegraphics[angle=0,width=140mm]{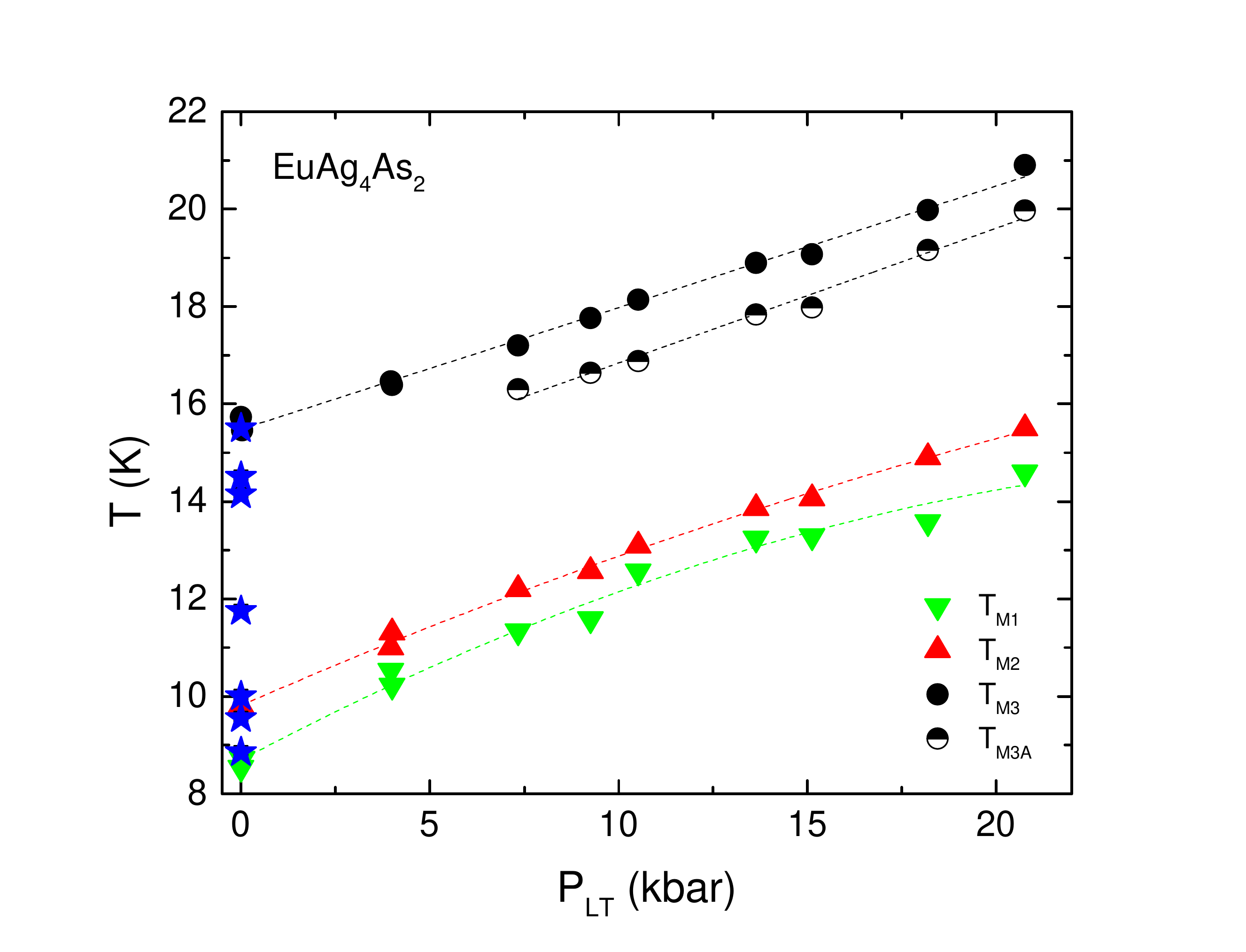}
\end{center}
\caption{(color online) Pressure dependence of the magnetic ordering temperatures in  EuAg$_4$As$_2$ determined from $\rho_{ab}(T)$ measurements (circles and triangles). Stars - ambient pressure transitions from heats capacity data.} \label{F6}
\end{figure}

The low temperature resistivity for two representative pressures, 0.02 kbar  and 18.2 kbar are shown in figures \ref{F5} (a),(b), respectively. (Note that the 0.02 kbar data are very close to the ambient pressure data taken outside the pressure cell, that suggests that for measurements protocol used the thermal mass of the pressure cell has no significant effect on the size of hysteresis.) At higher pressure the magnetic transitions are sharper and the hysteresis is larger. It should be noted that the upper magnetic transition ($T_{M3}$) that appears as a single anomaly at low pressures (Fig. \ref{F5}(c)) appears as two anomalies at higher pressures  (Fig. \ref{F5}(d)) indicating that either it splits under pressure or one of the neighboring, ambient pressure transitions becomes resolvable as pressure is increased (Fig. \ref{F6}).

Figure \ref{F6}  presents the pressure dependence of the three magnetic transitions as resolved by in-plane resistivity measurements. The two lower transition temperatures, $T_{M1}$ and $T_{M2}$, increase under pressure, the higher one, $T_{M3}$, splits into two ($T_{M3}$ and $T_{M3A}$) at and above $\sim 7.3$ kbar. Whereas $T_{M3}$ and  $T_{M3A}$,  increase linearly under pressure with the pressure derivatives $dT_{M3}/dP = 0.25$ K/kbar and  $dT_{M3A}/dP = 0.28$ K/kbar, for the lower transitions, $T_{M1}$ and  $T_{M2}$, the behavior under pressure is better fit with the second order polynomial with the initial ($P \to 0$) values of the derivatives $dT_{M1}/dP = 0.41$ K/kbar and  $dT_{M2}/dP = 0.34$ K/kbar.

\begin{figure}
\begin{center}
\includegraphics[angle=0,width=140mm]{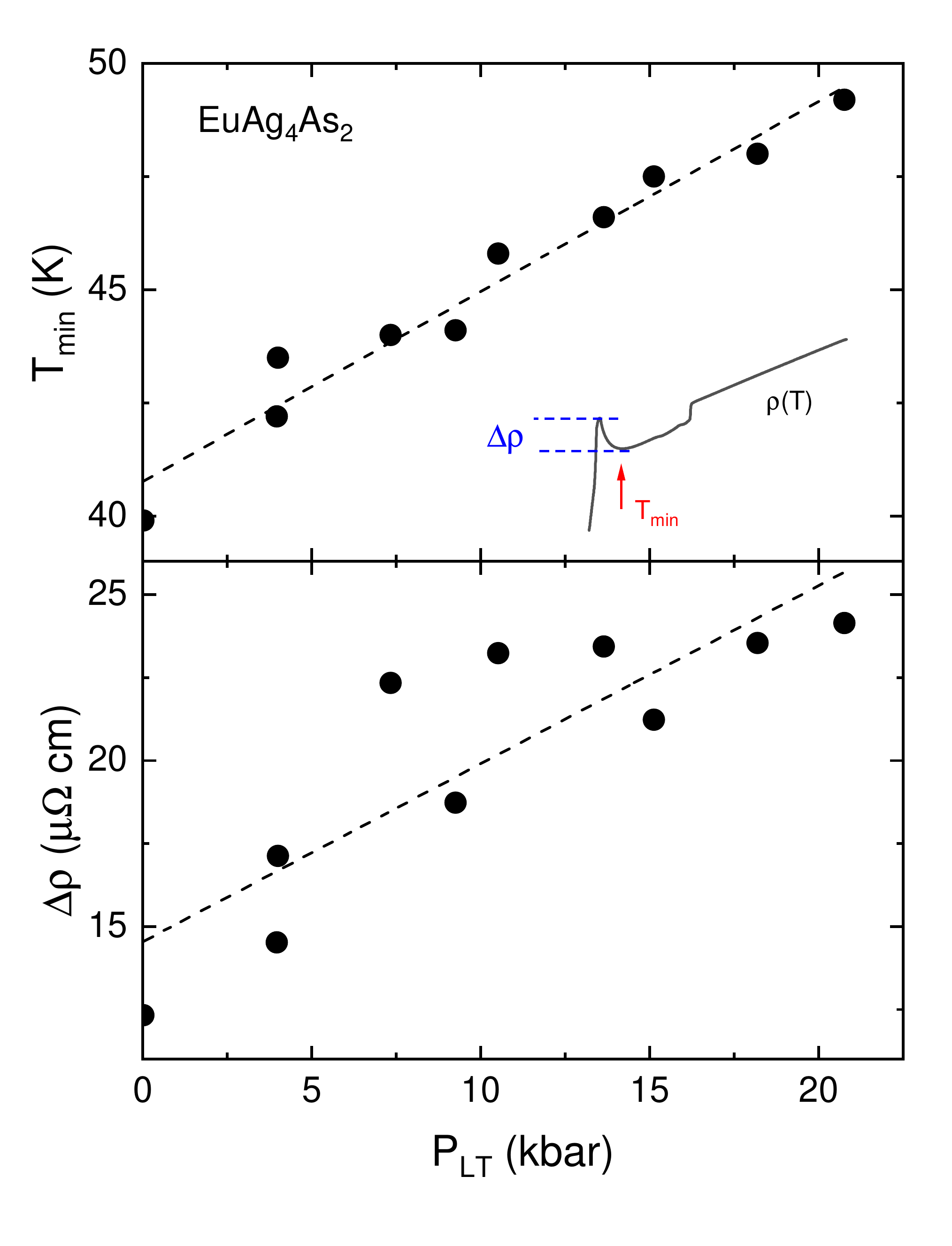}
\end{center}
\caption{(color online) Pressure dependence of the temperature of the resistivity minimum, $T_{min}$  (top panel) and the size of the resistivity upturn, $\Delta \rho$, (bottom panel) for in-plane resistivity of  EuAg$_4$As$_2$. Dashed lines are linear fits to the data.  The definitions of  $T_{min}$ and  $\Delta \rho$ are shown on the sketch in the top panel.} \label{F7}
\end{figure}

The resistivity minimum that precedes these magnetic transitions (Fig. \ref{F7}) shifts to higher temperatures under pressure at the rate of 0.42 K/kbar, which is faster than the upper magnetic transition. The size of the resistivity upturn also increases under pressure  (Fig. \ref{F7}). At least in part this is simply due to the increase of the temperature range over which the upturn is observed.

\begin{figure}
\begin{center}
\includegraphics[angle=0,width=140mm]{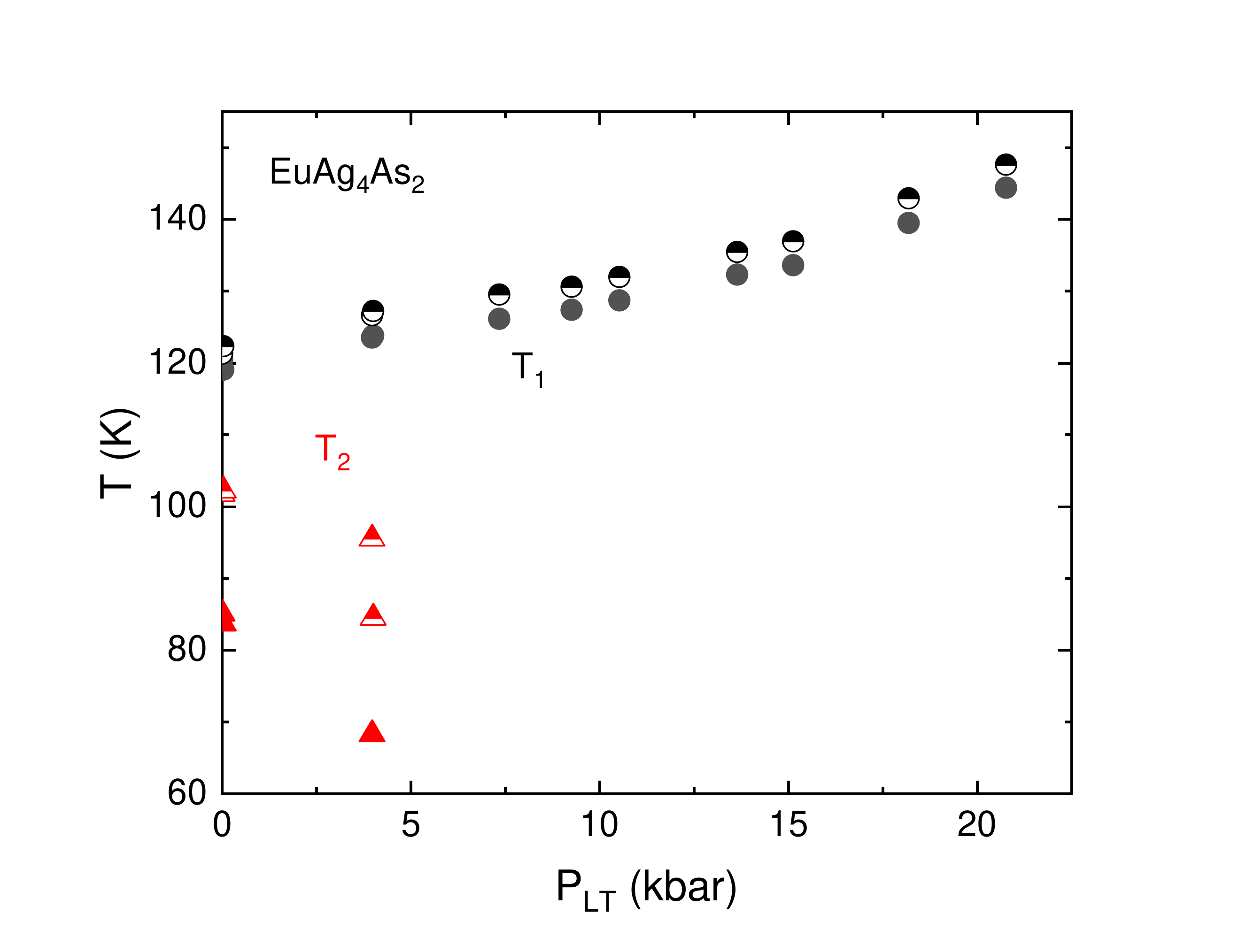}
\end{center}
\caption{(color online) Pressure dependence of the structural transitions temperatures, $T_1$ and $T_2$. Filled symbols - measured on cooling, half-filled - on warming   Data at $P \approx 4.0$ kbar were taken twice: on increase and release of pressure. } \label{F8}
\end{figure}

Now we turn to structural phase transitions under pressure. As seen in figures \ref{F4} and \ref{F8}, the  higher structural transition, $T_1$, is observed in all pressure range of this work. The transition temperature increases under pressure. The behavior clearly changes between two pressure ranges: below $\sim 14$ kbar the $T_1(P)$ is linear, with $dT_1/dP \approx 0.9$ K/kbar, whereas above $\sim 14$ kbar the pressure derivative changes by factor of 2, to  $dT_1/dP \approx 1.9$ K/kbar. The thermal hysteresis, $\sim 3.3$ K is basically pressure independent. Lower structural transition, $T_2$, initially decreases under pressure with the rate of in-between  - 3 and -4 K/kbar. Its signature in $\rho_{ab}(T)$, even at ambient pressure, is rather subtle and we are not able to detect it any more at and above $\sim 7.3$ kbar, possibly because rather strong, non-monotonic background. 

\begin{figure}
\begin{center}
\includegraphics[angle=0,width=140mm]{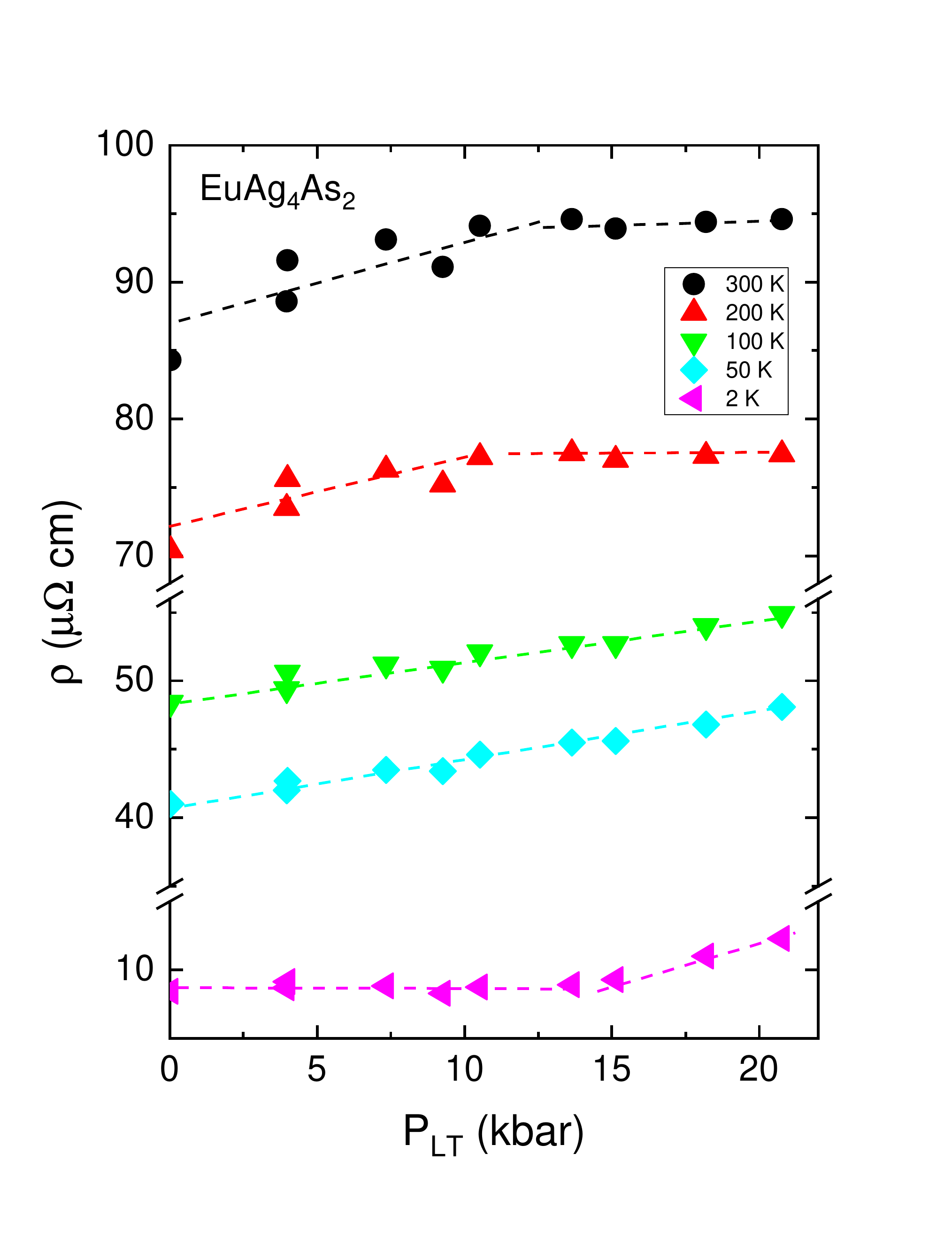}
\end{center}
\caption{(color online) Pressure dependence of  in-plane resistivity of  EuAg$_4$As$_2$ at selected temperatures. Dashed lines are guide for the eye.} \label{F10}
\end{figure}

It is noteworthy, that a weak, but discernible, anomaly in the 10-15 kbar range is also observed in pressure dependencies of the high and low temperature resistivity data, both above and below all noted transition temperatures. (Fig. \ref{F10}).

\begin{figure}
\begin{center}
\includegraphics[angle=0,width=140mm]{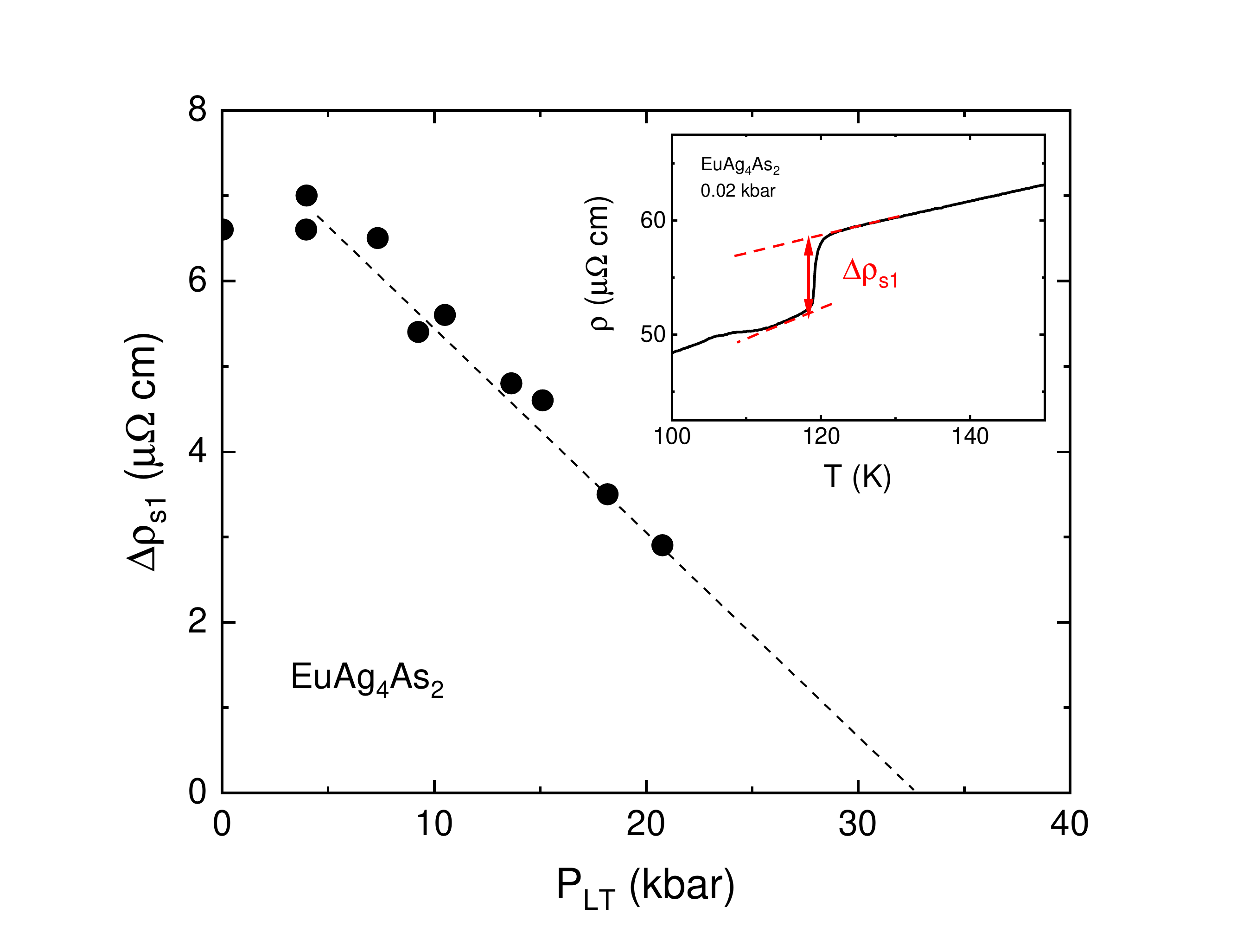}
\end{center}
\caption{(color online) Pressure dependence of the size of the resistivity upturn, $\Delta \rho_{s1}$, at the higher structural transition, $T_1$.  Inset shows definition of  $\Delta \rho_{s1}$ .} \label{F9}
\end{figure}

Finally, we can analyze the change the jump in resistivity at the upper structural transition as a function of pressure (Fig. \ref{F9}). Its value decreases under pressure with linear extrapolation to $\Delta \rho_{s1} = 0$ at $\sim 32.5$ kbar, that would correspond the (extrapolated) value of $T_1 \sim 165-170$ K. There are several possible scenarios of what might happen above $\sim 32.5$ kbar. Most probably, the structural transition will continue to exist with the transition temperature increasing further under pressure, however either without discernible feature in $\rho_{ab}$, or with inversion of such feature (decrease of $\rho_{ab}$ at $T_1$ on warming).

\section{Discussion and Summary}

Our resistivity measurements under pressure as well as ambient pressure thermodynamic and transport measurements further underscore the complexity of EuAg$_4$As$_2$ as a host for a multitude of structural and magnetic phases. In addition to documenting the evolution of different phases under pressure this work provides a roadmap for further studies, if undertaken.

\begin{figure}
\begin{center}
\includegraphics[angle=0,width=140mm]{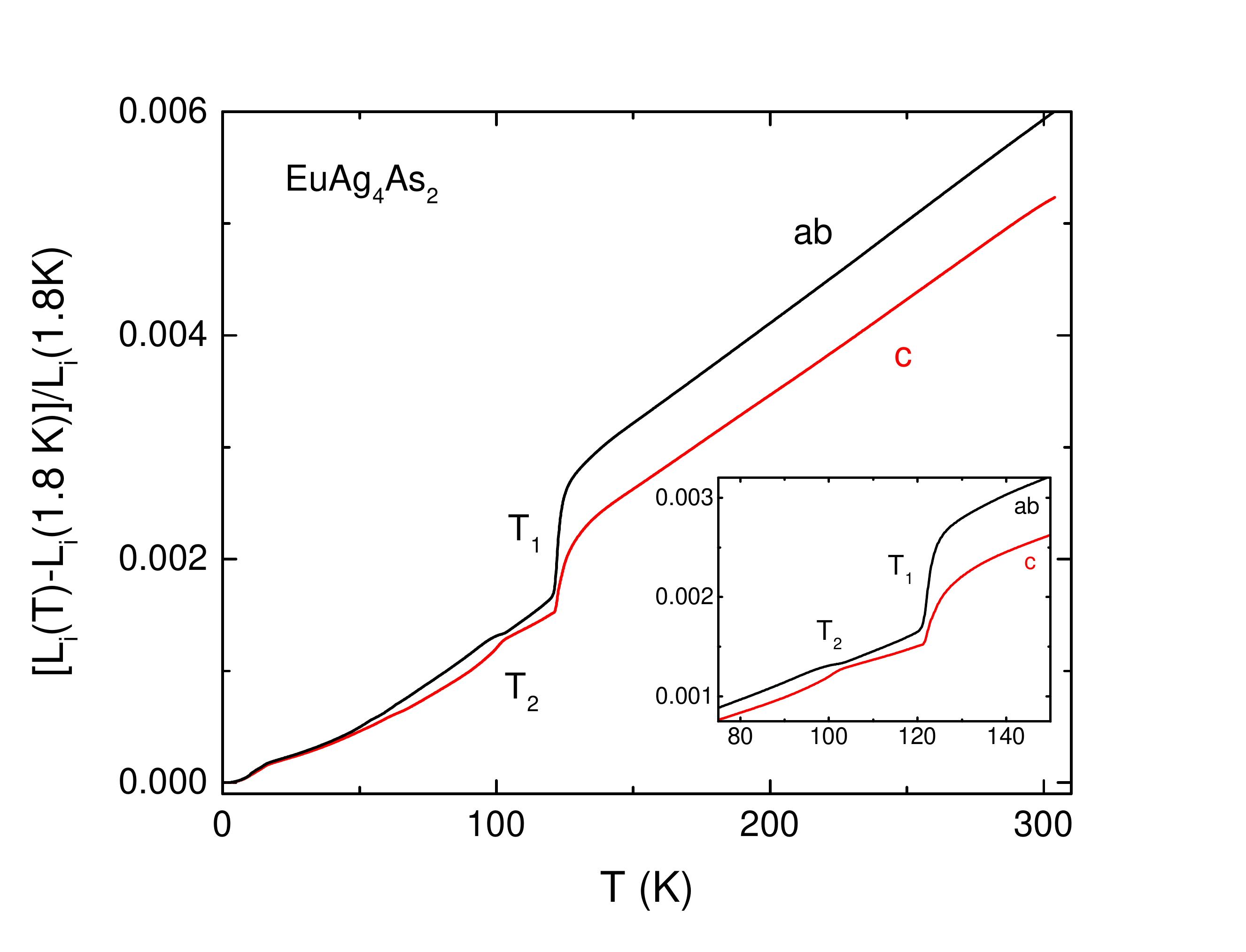}
\end{center}
\caption{(color online) Anisotropic, normalized to 1.8 K,  thermal expansion of EuAg$_4$As$_2$ measured on warming. Inset: enlarged region of two structural transitions in  anisotropic thermal expansion.} \label{FN4}
\end{figure}

The signs of the initial pressure derivatives of the structural and magnetic phase transitions are consistent with those inferred from the combination of the specific heat data  and the thermal expansion data in this work via Clausius - Clapeyron (1st order phase transitions) or Ehrenfest (2nd order phase transitions) relations. \cite{bar99a}  It is curious though that given the anisotropic, trigonal structure of EuAg$_4$As$_2$, its  in-plane and $c$-axis thermal expansion evolution is similar for the higher structural ($T_1$) and all three magnetic transitions, and only at lower structural transition $T_2$ different signs of thermal expansion are observed (Fig. \ref{FN4}). 

The resistivity of EuAg$_4$As$_2$ over wide temperature ranges  increases under pressure, suggesting either pressure - induced decrease of the density of states at the Fermi level, or decrease of mobility (increase of effective mass). Note that for the related compound, SrAg$_4$As$_2$, the opposite trend, a decrease of the in-plane resistivity under pressure was reported. \cite{she18b} 

The magnetic transitions shift to higher temperatures under pressure. The values of the pressure derivatives are rather conventional. 

The nature of the lower structural transition, $T_2$, and its evolution under pressure above $\sim 4$ kbar would require careful scattering studies to better understand. It is of interest, assuming this transition can be suppressed down to $T = 0$ K, if it has, even subtle, effect on the magnetic transitions.

The upper structural transition temperature, $T_1$, increases under pressure with a kink in $T_1(P)$ at $\sim 14$ kbar, whereas the jump in $\rho_{ab}$ associated with it gets smaller. Linear extrapolation suggests that the jump will disappear at 30-35 kbar. This behavior is very different from that of the structural transition in SrAg$_4$As$_2$ \cite{she18b} that has a minimum at $\sim 7.5$ kbar both in $T_s(P)$ and in $\Delta \rho_{xx}(P)$. Further studies are required to understand what will happen with the upper structural transition and its signature in the in-plane resistivity, and, broadly speaking, to the electronic structure, above $\sim 32.5$ kbar. 

The anomaly in the higher temperature, structural phase transition, $T_1(P)$, behavior  in the 10-15 kbar range  is also present in the $\rho(P)$ data  both above and below all noted transition temperatures. Are these anomalies associated with changes in the electronic structure and what is the origin of these changes?  This question will require further studies. Additionally, some understanding of the nature of the differences in pressure response of EuAg$_4$As$_2$ and SrAg$_4$As$_2$ would be desirable.

\begin{acknowledgments}

The authors thank Dominic Ryan and Elena Gati for useful discussions. Work at the Ames Laboratory was supported by the U.S. Department of Energy, Office of Science, Basic Energy Sciences, Materials Sciences and Engineering Division. The Ames Laboratory is operated for the U.S. Department of Energy by Iowa State University under contract No. DE-AC02-07CH11358. L.X. was supported, in part, by the W. M. Keck Foundation. Work at UCLA was supported by the U.S. Department of Energy, Office of Science, Office of Basic Energy Sciences under Award Number DE-SC0011978. 

\end{acknowledgments}

\clearpage

\clearpage

\clearpage

\clearpage

\clearpage

\clearpage

\clearpage

\clearpage

\clearpage

\clearpage

\clearpage

\end{document}